\documentstyle{article}
\renewcommand{\baselinestretch}{2}
\begin{document}
\begin{center}
{\Large\bf S U N}\\
\end{center}
{\large\bf H.J. HAUBOLD}, {\large Office for Outer Space,
United
Nations,\\
New York, NY 10017, USA}\par
\bigskip
\noindent
{\large\bf A.M. MATHAI}, {\large Department of Mathematics
and\\
Statistics, McGill University, Montreal, P.Q., Canada H3A
2K6}\par
\bigskip
\noindent
{\Large\bf SUN}\\
\medskip
\noindent
\hspace{1cm}
{\Large\bf Solar Structure}\par
\hspace{1.5cm}\underline{\large\bf Internal Structure}\par
\hspace{2.0cm}core\par
\hspace{2.0cm}radiative zone\par
\hspace{2.0cm}convective envelope\par
\hspace{2.0cm}photosphere\par
\hspace{2.0cm}chromosphere\par
\hspace{2.0cm}corona\par
\hspace{1.5cm}\underline{\large\bf Solar Activity}\par
\hspace{2.0cm}granules\par
\hspace{2.0cm}spicules\par
\hspace{2.0cm}sunspots\par
\hspace{2.0cm}prominences\par
\hspace{2.0cm}solar flares\par
\clearpage
\hspace{1cm}{\Large\bf Solar Rotation}\par
\hspace{1cm} {\Large\bf Solar Magnetic Field}\par
\hspace{1cm} {\Large\bf Solar Thermonuclear Energy
Generation}\par
\hspace{1cm} {\Large\bf Solar Evolution}\par
\hspace{1.5cm} \underline{\large\bf Presolar Evolution
Stages}\par
\hspace{2cm}cloud\par
\hspace{2cm}globule\par
\hspace{2cm}protostar\par
\hspace{2cm}sun\par
\hspace{1.5cm}\underline{\large\bf Postsolar Evolution
Stages}\par
\hspace{2cm}red giant\par
\hspace{2cm}white dwarf\par
\hspace{2cm}black dwarf\par
\bigskip
\begin{center}
{\Large\bf References}
\end{center}
\clearpage
{\Large\bf SUN}
\bigskip

The Sun is a main-sequence star, one of over 100 billion stars
in
the Milky Way Galaxy. It takes the Sun over
200 million years to complete one orbit of the galaxy. It is
located at present close to the Sagittarius-Carina
spiral arm, in what is called the Orion spur.\par
The orbital trajectory is not planar, but undulating, the Sun
passing through the principal plane of the galaxy once every
$\sim$
30 million years.
The planets of the Solar System revolve about the Sun in a
plane normal to the Sun's circum-galactic trajectory.
Thermonuclear
reactions in the Sun's core convert
hydrogen into helium, so that the helium fraction is
progressing
increasing through time. The reactions produce the energy to
make
the Sun a yellow-orange main-sequence star
of spectral type dG2 (d indicates dwarf). The central
temperature
is about 15 million K; the surface temperature is about 6000
K.\par
The Sun is a nearly-perfect sphere of gas with a diameter of
1,400,000
km,
held together by its own gravity. Its dimensions are at
present not
static
and recent findings have shown that the Sun's diameter is at
the
present time decreasing about one meter every hour. This
decrease
may be
a long-term oscillation which may be one effect of the
long-term stabilization of the Sun's output of energy. It has
also
been
found that every 2h 40min the Sun's surface pulses at a speed
of
6 $kmh^{-1}$. Thus its surface moves in and out to
change the diameter by nearly 10 km.\par
The Sun contains 99.86\% of all matter in the solar system. It
is
the source of light and heat for all planets and the support
of
life on Earth. A
number of properties of the Sun is listed in Table 1-1 (cp.
also
Allen 1973; Lang 1980).\par
\bigskip
\noindent
{\Large\bf Solar Structure}\\
{\large\bf Internal structure.}
The structure of the Sun is determinated by the conditions of
mass
conservation, momentum conservation, energy conservation, and
the
mode of energy transport. The Sun is an oblate spheroid, like
all
the major bodies in the solar system, but in a first
simplifying
approach to describe the solar
structure, the effects of rotation  and magnetic fields will
be
neglected here so that
the Sun is taken to be spherically symmetrical. Calculating a
solar
model means the determination of pressure, temperature and
chemical
composition as a function of mass or radius through the Sun,
(Chandrasekhar 1967; Kourganoff 1973).\par
Two forces keep the Sun in hydrostatic equilibrium in its
current
stage of evolution: the gravitational force directed inward
and the
total pressure force directed outward. The equation of
hydrostatic
equilibrium is
\begin{equation}
\frac{dP}{dr}=-\rho \frac{GM_r}{r^2},
\end{equation}
where P is the pressure, r the radial distance from the
center,
$M_r$ the mass within a sphere of radius r, $\rho$ the matter
density, and G the gravitational constant. This equation is
consistent with radius changes, but requires the kinetic
energy
involved in
expansion or contraction of the solar body to be small
compared to
the gravitational potential of the Sun. For an order of
magnitude estimate, equation (1) can be written
\begin{equation}
\frac{dP}{dr}\approx \frac{P_c-P_o}{R_\odot}\approx
\frac{P_c}{R_\odot} \approx \frac{GM_\odot
\bar{\rho}}{R_\odot^2},
\end{equation}
where $R_\odot$ is the solar radius, $M_\odot$ the solar mass,
$\bar{\rho}$ the mean matter density of the solar gas sphere,
$P_c$
is the central and $P_o$ the surface pressure respectively,
where
the latter can be neglected.\par
The equation of mass conservation
\begin{equation}
\frac{dM_r}{dr}=4\pi r^2\rho,
\end{equation}
constrains the integral of the density over the volume to be
equal
to the mass and leads to the estimation of the mean matter
\underline{density} for
the
Sun
\begin{equation}
\frac{M_\odot}{R_\odot} \approx R_\odot^2
\bar{\rho}\Longrightarrow
\bar{\rho}_\odot \propto \frac{M_\odot}{R_\odot^3},
\end{equation}
where the symbol $\propto$ means ``varies as''. In the general
case, $\rho \propto M/R^3$, the constant of proportionality
depends
on the radial mass distribution and the radial distance R
(Schwarzschild 1958; Haubold
and Mathai 1987, 1992).
Using eq.(4) and eq.(2), the central \underline{pressure} of
the
Sun can be
estimated to
be
\begin{equation}
P_c \propto G\frac{M_\odot^2}{R_\odot^4}.
\end{equation}\par
In the general case, $P\propto GM^2/R^4$, the constant of
proportionality is determined by the radial distribution of
mass in
the Sun, and the particular radial distance R at which P is
measured (Schwarzschild 1958; Haubold and Mathai 1987,
1992).\par
The interior of the Sun is entirely gaseous and the great
majority
of atoms are stripped of their
electrons. The solar gas behaves under these physical
conditions
nearly like
a perfect gas, governed by the ``equation of state''
\begin{equation}
P=\frac{k}{\mu m_p}\rho T,
\end{equation}
where $m_p$ is the mass of the proton, $k$ is Boltzmann's
constant,
and
$\mu$ is the mean molecular weight. This equation of state
relates
the pressure, temperature, density and chemical composition,
and is
related to other
thermodynamic quantities. Then the central
\underline{temperature} of
the Sun can be estimated from the perfect gas law in eq.(6),
that
is
\begin{equation}
T_c\approx\frac{\mu
m_p}{k}\frac{P_c}{\bar{\rho}}\Longrightarrow
T_c \propto \frac{
m_pG}{k}\mu \frac{M_\odot}{R_\odot}.
\end{equation}
This formula determines the temperature at the centre of the
Sun
according to its mass, radius and mean molecular weight of the
solar matter. In the general case, $T\propto \mu M/R$, the
constant
of proportionality depends on the mass distribution and the
radial
distance R (Schwarzschild 1958; Haubold and Mathai 1987,
1992).
\par
When X, Y, and Z are the mass fractions of hydrogen, helium,
and
heavy elements, respectively, then it holds by definition
X+Y+Z=1.
The mean molecular weight $\mu$ in eq.(6) can be calculated
when
the degree of ionization of each chemical element of solar
matter
has been specified. For solar gas composed of fully ionized
hydrogen, there are two particles for every proton and it is
$\mu
= 1/2$. For a gas composed of fully ionized helium it is $\mu
=
4/3$. For all elements heavier than helium, usually referred
to by
astronomers
as metals, it holds that their atomic weights are
twice their charge and accordingly $\mu =2$. Thus the mean
atomic
weight
for fully ionized gas is
\begin{equation}
\mu=\frac{1}{2X+(3/4)Y+(1/2)Z}.
\end{equation}
The solar matter is at present approximately 75\% hydrogen,
23\%
helium and
2\% metals by mass fraction. Throughout the solar interior,
$\mu_\odot$ is approximately 0.59, except at the surface,
where
hydrogen and helium are not fully ionized, and in the core,
where
the chemical composition is altering due to nuclear burning
(cp.
Table 1-2) (Kavanagh 1972; Bahcall 1989).\par
An equation of continuity must also be satisfied by the
radiation
\begin{equation}
\frac{dE}{dt}+\frac{dL_r}{dr}=0,
\end{equation}
where $dE/dt$ is the rate of energy production per unit
thickness
of the shell of radius r.
The equation of \underline{energy conservation} is
\begin{equation}
\frac{dL_r}{dr}=4\pi r^2\rho\epsilon
=-\frac{dE}{dt},
\end{equation}
where $L_r$ denotes the total net energy flux through a
spherical
shell of radius $r$ and $\epsilon$ is the net release of
energy per
gram per second by thermonuclear reactions occurring in the
gravitationally stabilized solar fusion reactor. It is assumed
in
eq.(9) that the energy produced by nuclear reactions equals
the
photon luminosity of the Sun, thus neglecting gravitational
contraction and subtracting energy loss through neutrino
emission.
The mean
energy
generation rate for the Sun can be inferred from eq.(10), that
is
\begin{equation}
\bar{\epsilon}\approx\frac{L_\odot}{M_\odot}.
\end{equation}\par
Finally the thermonuclear energy produced in the solar core is
transported by radiation through the solar body to the
surface. The
force due to the gradient of the \underline{radiation
pressure}
equals the
momentum absorbed from the radiation streaming through the gas
\begin{equation}
\frac{dP}{dr}=-\frac{\kappa\rho}{c}\frac{L_r}{4\pi r^2},
\end{equation}
where $P=aT^4/3$ is the radiation pressure, $\kappa$ is the
opacity
of the solar matter, $1/\kappa \rho$ is the
mean free path of photons, and c is the velocity of light.
The coefficient of the radiation density, $a$, is related to
the
Stefan-Boltzmann constant $\sigma$ since $\sigma = ac/4$.
Equation(12) is the energy transport equation taking into
account
the fact that energy transport in the deep interior of the Sun
is
exclusively managed by
radiation. From eq.(11) follows the temperature gradient
driving
the radiation flux, that is
\begin{equation}
\frac{dT}{dr}=-\frac{3}{4ac}\frac{\kappa
\rho}{T^3}\frac{L_r}{4\pi
 r^2},
\end{equation}
allowing an estimate of the solar luminosity
\begin{equation}
\frac{T_c}{R_\odot} \approx \frac{1}{ac}\frac{\kappa
\bar{\rho}}{T_c^3}\frac{L_\odot}{R_\odot^2} \Longrightarrow
L_\odot
\propto ac(\frac{Gm_p}{k})^4\frac{\mu^4}{\kappa}M_\odot ^3,
\end{equation}
taking into account eqs.(4) and (7). The luminosity is
independent
of the radius; it depends on the opacity and increases with
mass. Eq.(14) is an important result of the theory of the
internal
structure of solar-type stars, called theoretical
mass-luminosity
relationship. The fundamental result as given by eq. (14) is
that
the luminosity of the star is simply determined by its mass,
since
this rule is based on the fact that the transfer of energy
from the
stellar interior towards the surface is managed by radiation.
The stellar energy sources must somehow adapt
to the stellar opacity. The luminosity of a solar-type star is
determined largely by photon opacity and not by the energy
source.\par
Gamma-ray photons produced in thermonuclear reactions in the
core
of the Sun are being scattered, absorbed or re-emitted by free
electrons,
ions, and atoms on their way to the surface of the Sun. The
\underline{opacity}
$\kappa$ in eq.(12) is the measure of the solar material's
efficiency at inhibiting the passage of the photons through
the
solar interior. The actual value of the opacity depends on
various
processes which may operate simultaneously: bound-bound
transitions, bound-free transitions, free-free transitions,
and
scattering of photons by free electrons, ions and atoms.
Scattering
of photons by
free electrons is the most important process for the solar
core.
Approaching the solar surface, bound-free transitions take
over to
determine the opacity of solar matter. The structure of the
Sun
depends in a sensitive way on the opacity, for if $\kappa$
changes,
the Sun must readjust all its parameters to allow the energy
generated in the core to stream to the surface, not being
blocked
at
any point in the solar interior.\par
Boundary conditions for the system of nonlinear differential
equations (eqs.(1), (3), (10), (12)) have to be specified to
arrive
at
specific solutions: At the solar centre it is $r = 0, M_r = 0,
L_r
= 0$,
and at
the assumed solar surface (this is actually the photosphere)
it
holds $M_r=M_\odot$, and for an age of $t_\odot
=4.5\times10^9$ years, $r=R_\odot$, $L=L_\odot$.
Mass, radius, surface temperature, surface chemical
composition,
and luminosity of the Sun are known by observation. Using the
conservation laws and known properties of gases (equation of
state,
opacity, energy generation rates), the internal structure of
the
Sun can be calculated in matching the observed properties at
the
solar surface. However, because the equations of solar
structure
form a system of first-order nonlinear simultaneous
differential
equations, they have to be integrated numerically to obtain a
very
detailed picture of the run of physical variables throughout
the
Sun. Order of  magnitude estimations provided in
eqs.(4),(5),(7),
(11) and (14) can be considered only to be a first approach to
the
problem (Mathai and Haubold 1988).
Figure 1-2 shows the numerical results of a standard solar
model
based on the system of differential equations as described
above (Sears 1964; Sackmann, Boothrayd, and Fouler 1990;
Guenther
et al. 1992).\par
Chemical composition changes with time (cp. eq. (10)) due to
thermonuclear
reactions in the solar core that results in a continuously
evolving structure, the calculation of which adds another
system
of differential equations (kinetic equations) to the set of
differential equations described above (Schwarzschild 1958;
Kourganoff 1973).\\
\underline{core}
The core of the Sun is a gravitationally stabilized fusion
reactor.
There, energy is produced by conversion of hydrogen into
helium.
Each hydrogen atom weighs 1.0078 atomic units and each helium
atom
is made from four hydrogen atoms thus weighing 4.003 atomic
units.
Accordingly,
the difference of 0.0282 atomic units, or 0.7\% of the mass
$m$, is
converted into energy $E$ according to Einstein's formula
$E=mc^2$,
where $c$ is the velocity of light. Most atoms in the core of
the
Sun
are
entirely stripped of their electrons by the high temperature
and
opacity is governed by scattering
of photons by free electrons, by inverse bremsstrahlung on
ionized
hydrogen and helium, and by bound-free scattering by elements
heavier than helium.\\
\underline{radiative zone}
The radiative zone is a region of highly ionized gas. There
the
energy transport is primarily by photon diffusion and is
described
in terms of the Rosseland mean opacity (this is a weighted
inverse
mean of the opacity over all frequencies, which can be used
when
the optical depth is very large and radiative transport
reduces to
a diffusion process).\\
\underline{convective zone}
In the outer regions, atoms may keep their electrons because
of the
low temperature and ions and even neutral hydrogen exist. Here
many
atomic absorption processes occur, mainly bound-free
transitions.
The high opacity makes it
difficult for photon radiation to continue outward and steep
temperature gradients are established which lead to convective
currents. The outer envelope of the Sun is in convective
equilibrium. It is the location where sunspots and other solar
activity phenomena are generated. Observationally, the outer
solar
atmosphere following the convective zone has been divided into
three spherically symmetric layers - the photosphere,
chromosphere,
and corona - lying successively above one another (Zirin
1988).\\
\underline{photosphere}
The outer limit of the photosphere is the boundary of the
visible
solar disk as seen in white light. Most of the
radiation emitted by the Sun originates in the photosphere,
which
is only
about 500 km thick. This radiation is in equilibrium and the
Stefan-Boltzmann law can be applied to
calculate the effective temperature of the solar photosphere,
which
is $T_e=5780 K$. According to the Stefan-Boltzmann law each
square
centimeter of the solar surface having the temperature T
emits, in
all directions, light of $\sigma T^4$ ergs per second.
Subsequently, the total emission of the Sun in one second,
i.e. the
luminosity, equals
\begin{equation}
L_\odot=4\pi R_\odot ^2 \sigma T_e^4.
\end{equation}
This fundamental relation also determines the radius of the
Sun
when its
luminosity and surface temperature are known.
The spectrum of the photosphere consists of absorption
lines superimposed on an approximately blackbody continuum.\\
\underline{chromosphere}
A thin transition region extending 5000 km above the
photosphere is
called the chromosphere.
Considerably hotter than the photosphere, the
chromosphere is heated by
hydromagnetic waves and compression waves originated by
spicules
and granules. The temperature of the chromosphere is about
10,000
K and it has an emission spectrum.\\
\underline{corona}
During a total solar eclipse the outermost atmosphere of the
Sun
can
be seen. Called the solar corona (q.v.), this is a hot gas
merging
gradually into the transparent interplanetary medium, and
flowing outward from it
is the \underline{solar wind}. Current theories indicate that
the
corona is heated by the dissipation of mechanical energy
stemming
from the convection zone, or by dissipation of magnetic energy
by
field-line reconnection. The kinetic temperature of the solar
corona is
about $2\times10^6 K$ and its gas has a density of about
$10^{-15}
 gcm^{-3}$. Solar x-ray radiation originates in the corona.\\
{\large\bf Solar activity.}
The Sun emits radiation in a wide range of the energy spectrum

from long radio waves (300m) to x-rays (0.1nm), including
high-energy particles (\underline{cosmic rays} q.v.). Almost
95\%
of the radiated
energy
is concentrated in a relatively narrow band between 250 nm and
2500
nm. The total radiation received from the Sun is called the
\underline{solar
constant}; it was formerly regarded as a fixed value, $2.00
\pm
0.04\; calcm^{-2}min^{-1}$, alternatively $1.36 \times 10^6
ergs cm^{-2}s^{-1}$ (although difficult to measure), but from
satellite
observations it is now confirmed as a variable (up to about
0.5\%) (Herman and Goldberg 1978; Sofia 1981; Schatten and
Arking
1990). The transient phenomena occurring in the solar
atmosphere
can be grouped together under the term solar activity:
sunspots and
faculae occur in the photosphere; flares and plages belong
to the chromosphere; and prominences and coronal structures
develop
in the corona. All solar activity phenomena are
connected in this way or another with the 11 and 22-year
sunspot
cycle.\\
\underline{granules}
Granules are huge convective cells of hot gases, 400-1000 km
in
diameter,
spread in a cellular pattern over the entire
photosphere except at sunspots. Granulation supports the
transfer
of energy
from the convective zone outward into space. Granules behave
as
short-lived bubbles, lasting only 3 to 10 minutes, that rise
and fall at a velocity of about $0.5\; kms^{-1}$, thereby
moving
vertically a distance of the order of 200 km.\\
\underline{spicules}
Spicules look like hairs of gas rising and falling at the
upper
chromosphere, reaching into the corona. They last as long as
10
minutes, attaining vertical speeds of up to $20\; kms^{-1}$
getting
upward
to as high as 15,000 km. Spicules array themselves into
chromospheric networks
establishing giant supergranulated cells with gases rising in
the
center and descending at their outer boundaries.\\
\underline{sunspots}
Sunspots are relatively cool and dark markings on the Sun's
photosphere, which exhibit distinct cycles. They are
concentrations
of strong magnetic fields (2000-3000 Gauss), with diameters
less
than about 50,000 km and lifetime of a few days to weeks. A
sunspot
generally
develops
a very dark central region, called the umbra, which is
surrounded
by the penumbra. The 11-yr sunspot cycle consists of
variations in
the
size, number, and position of the sunspots (Fairbridge 1987a).
It
is
extremely
variable in length (actually, 7 to 17 yr.), the high-activity
cycles (to $\geq 200$ spots)
are generally short
(9-10
yr), and the low activity cycles (sometimes $\leq 50$ spots)
are
long
(12-13 yr). In the sunspot cycle
the number of sunspots usually peaks 2-3 yr after the
beginning of
each
cycle and decays gradually, but low activity cycles may have a
reversed symmetry.  First spots of the cycle appear at
higher latitudes, mostly between $20^\circ$ and $35^\circ$,
and as
the
spots increase in size and number they occur closer to the
equator.
Very few spots are observed outside the latitude range of
$5^\circ
- 35^\circ.$ The magnetic polarity of
the sunspot groups reverses in each successive cycle so that
the
complete cycle lasts 22 years, the so-called ``Hale cycle''.
Recent
observations have indicated that the
magnetic solar cycle is a coherent phenomenon throughout the
solar surface. For each pair of 11-yr cycles, the one with
north or
leading magnetic orientation is usually stronger than the
south
one. Between 1645 and 1715 very few sunspots were seen,
a time period called Maunder minimum. This period was
associated
with a long cold spell in Europe, known as Little Ice Age.
Carbon-14 measurements from tree rings and Beryllium-10
measurements
from arctic ice-cores confirm the low solar
activity level
at that time (Sonett 1984; Beer 1987, Fairbridge 1987b). The
solar
11-year cycle has
been recorded on a
regular
basis since the beginning of the 18th century classified by
Wolf's
quantity $N$ of the number of sunspots $N_s$ plus ten times
the
number $G$
of sunspot groups: $N=W(N_s+10G)$, where $W$ is a weighting
factor assigned to an individual observer to account for
variation
in equipment, atmosphere conditions, and observer enthusiasm
(Gibson 1973).
The quantity $N$ is widely used as an indicator of sunspot
activity
and is commonly called the Zurich sunspot number. More
recently
Bracewell (1989) was able to show that the quantity
$N=\pm(N_s+10G)^{3/2}$, with $\pm$ denoting the dipole
orientation,
is nearly a sinusoidal function of time with a period of
$22.2\pm2$ years. Using proxy data for ancient sunspot periods
(such as auroral frequency), the average of the 11-yr cycle is
11.12 yr, and thus coupled to the magnetic period.\\
\underline{prominences} They are regions of cool $(10^4K)$,
high-density gas embedded in the hot $(10^6K)$, low-density
corona.
Prominences can be observed as flamelike tongues of gas that
appear
above the limb of the Sun when observed in the light of the
$H\alpha$ line. They occur in regions of horizontal magnetic
fields, because these fields support prominences against the
solar
gravity, and indicate the transition from one magnetic
polarity to
the opposite.\\
\underline{flares}
Sunspots are accompanied by large eruptions called solar
flares
emitting high-energy particles and radiation in a very broad
spectrum of energy. A solar
flare is actually the result of an intensely hot
electromagnetic
explosion in the corona and produces vast quantities of x-rays
which brighten the chromospheric gases. Typical
lifetimes of solar flares are one to two hours and the
temperature
in flares can reach several million degrees. Flare particles
ejected into outer space reach the Earth in a few hours or
days and
are the cause of disruptions in radio transmission. Aurora and
magnetic storms are due to strong solar flare eruptions. The
peak
of solar flare activity is lagged by the sunspot cycle,
usually 1-2
yr, but some high-energy eruptions may occur at any time ; the
mean
cycle of flare frequency is 0.417 yr.\par
\bigskip
\noindent
{\Large\bf Solar Rotation}\\
Solar rotation was first accurately measured in the last
century
(in 1863) by R.C. Carrington (q.v.) who used the position of
prominent spots as marker
points to determine a synodic period of about 27 days.
Beginning
from the first year of observation the solar rotations are
indentified by ``Carrington numbers''.
The solar surface, however, exhibits differential rotation, as
well
as a coherent pattern
of activity related to magnetic fields, and globally coherent
oscillation modes. All three phenomena can be employed to shed
light on the structure and dynamics of the Sun. Particularly
\underline{helioseismology}, the study of solar oscillation,
made
it possible
to measure the depth of the solar convection zone, the
internal
rotation profile, the sound speed throughout the Sun, and the
solar
helium abundance, (Deubner and Gough 1984; Hill and
Kroll 1992). Employing a standard model for
the internal structure of the Sun, it has been shown with
linear
adiabatic perturbation theory that small-amplitude
oscillations of
the solar body about its equilibrium state can be classified
into
three types:(i) pressure modes (p-modes), where the pressure
is the
dominant restoring force; (ii) gravity-modes (g-modes), where
gravity or buoyancy is the dominant restoring force; and a
class of
surface or interface modes
(f-modes), which are nearly compressionless surface waves. The
existence of all three modes has been confirmed by solar
observations. The solar
rotation rate through a large part of the solar interior has
been
estimated, utilizing for the most part observations of the
p-mode
frequency splittings. Each mode is
characterized by an eigenfunction with frequency eigenvalue
$\nu_{nlm}$, where n, l, and m are integer ''quantum''
numbers; n
counts the number of radial nodes in the wavefunction, and l
and m
describe the nodes in colatitude and longitude, respectively.
Rotation breaks the spherical symmetry of the Sun. Because of
that
the p-mode frequencies are not completely degenerate in m, and
the
frequencies $\nu_{nlm}$ in an nl-multiplet are said to be
split in
analogy to the Zeeman splitting of degenerate atomic energy
levels.
Because of observational limits it is not yet possible to
observe
values of splittings for individual $m$, to be used for
inversion.
However, results of observations are available in terms of
efficients $a_j(j\leq5)$ of least-squares fits of the
splittings
\begin{equation}
\nu_{nlm}-\nu_{nl0}=L\sum_ja_j(n,l)P_j^{(l)}(\frac{m}{L}),
\end{equation}
where $P_j^{(L)}$ is a polynomial of degree $j$ and
$L=(l[l+1])^{1/2}.$ The coefficients $a_j(n,l)$ of odd $j$
reveal
the information about the internal rotation of the Sun (Fig.
1-5).
The analysis of observational data reveal that the
latitude-dependent solar rotation profile as observed at the
solar
surface
extends down through the convective envelope. In the radiative
zone
the rotation seems to have a solid-body profile. (Schou,
Christensen-Dalsgaard, and Thompson 1992; Hill, Oglesby, and
Gu
1992). Todate there
exists no obvious theoretical explanation for this
helioseismologically inferred solar rotation profile.\par
Measurements of the individual frequencies of normal modes of
the
oscillating Sun may reveal the internal rotation profile. The
ultimate goal of helioseismology is, however, to use all
available
pulsation data, including growth rates, phases, different
modes -
and not just observed frequencies - to search the internal
structure
and evolution of the Sun. Those data will definitely
contribute to
improve the inadequate treatment of convective trasnport of
energy
in the envelope of the Sun by the mixing length theory as well
as
to solve the solar neutrino problem for the gravitationally
stabilized solar fusion reactor. Eigenmodes of pulsations of
different degree carry information of physical conditions in
quite
different parts of the Sun. High-degree modes (Figures 1-5 and
1-6)
are restricted to solar sub-surface layers, where solar
activity
phenomena have their origin. Contrary to this, low-degree
modes
(Figure 1-
7) propagate all the way through the solar body to the regions
where the solar neutrino flux is generated. Figures 1-5, 1-6,
and
1-7 are equatorial cross sections from a model of the
vibrating Sun
(Weiss and Schneider 1991).\par
According to observation and theory of stellar evolution,
young
stars rotate rapidly. If the central part of the Sun still
rotates
rapidly, this should lead to a small oblateness in the Sun's
disk,
about 1 part in $10^5$. The extreme observational values
reported
for the solar oblateness lie between $5.0 \pm
0.7\times10^{-5}$
(Dicke and Goldenberg 1967) and $9.6 \pm 6.5\times 10^{-6}$
(Hill
and Stebbins 1975), with a proposal that this quantity varies
with
the solar cycle (Dicke et al. 1987). The oblateness of the Sun
is
still a hotly debated issue in observational and theoretical
solar
physics.\par
\bigskip
\noindent
{\Large\bf Solar Magnetic Field}\par
\bigskip

All transient phenomena occurring in the solar atmosphere are
connected with magnetic fields leading to a
22-year Hale cycle. Todate all observed phenomena due to
subsurface solar magnetic fields are inferred from the laws of
magnetohydrodynamics. In sunspots the magnetic-field lines are
bundled and magnetic fields reach values of 2000 to 3000
Gauss. The mean magnetic-field intensity measurable at the
solar
surface is only approximately 1 Gauss. The small-scale
features of magnetic activity on the solar surface are
continously
changing with a degree of randomness as a result of
complicated
turbulent and ordered convective motions in the envelope of
the
Sun. The
large-scale sunspot cycle, however, shows a well-defined
behavior
as a result of convection and generation of poloidal and
toroidal
magnetic fields within the differentially rotating Sun. Near
the
base of the convection zone the magnetic field may reach an
amplitude of $10^5$ Gauss.\par
The existence and generation of magnetic fields in the deep
interior of the solar body is still a very controversial
issue. The
generally accepted view is that the convective envelope of the
Sun
is a converter of turbulence and differential rotation into an
oscillating magnetic toroid and dipole. The magnetic field is
confined to the convective envelope and is generated there by
a
dynamo mechanism, thereby consuming energy liberated by
thermonuclear reactions in the gravitationally stabilized
fusion
reactor of the Sun. Energy generated in the core of the Sun is
used
to drive convection and differential rotation in the envelope
of
the Sun. Dynamo models successfully explain the periodic
amplification of the solar magnetic field and the observed
butterfly diagram of sunspots, respectively. Almost all these
models rely on assumptions that employ stochastic mechanisms
for
the explanation of the 22-years solar activity cycle (Stix
1989).\par
Contrary to the stochastic approach to the generation of the
solar
magnetic field, it is possible in principle to explain the
magnetic field as the result of the collapse of the
primitive solar nebula. The radiative core of the Sun may have
conserved its primordial magnetic field, locked into matter.
It can
be supposed that the radiative core of the Sun has a high
electric
conductivity conserving its low-order magnetic multipoles.
Because
a magnetic dipole existing in a fluid conductor is unstable
towards a splitting along its symmetry planes and rotation
about
$180^\circ$, the dominant magnetic field in the core has a
quadrupole configuration. This quadrupole model for the solar
magnetic field could explain many of solar magnetic activity
phenomena, but has not yet been confirmed by observations
(Kundt
1992).\par
\bigskip
\noindent
{\Large\bf Solar Thermonuclear Energy Generation}\par
\bigskip

The Sun shines because of the process of fusion where four
protons
fuse to form an alpha
particle $\alpha$, two positrons $(e^+)$, and two neutrinos
$(\nu_e)$, that is, $4p \rightarrow \alpha + 2e^+ + 2\nu_e$.
In
this
fusion process of hydrogen nuclei into helium nuclei, the
latter
also known
as alpha particles, the fusion can be accomplished through two
different
series of principal reactions: 98.5\% of the energy generation
in
the present day Sun comes from the proton-proton chain (p-p
chain);
1.5\% of the solar energy output is due to the
Carbon-Nitrogen-Oxygen cycle (CNO cycle). The p-p chain and
the CNO
cycle are shown
in Table 1-3; there the third column indicates the percentage
of the solar terminations of the p-p chain in each reaction.
Since the dependence of the energy generation rate $\epsilon$
(cp.
eq. (8)) on the temperature is quite different between p-p
chain
and
CNO cycle, the p-p chain dominates at low temperature $(T\leq
18\times10^6\;K)$ (and the CNO
cycle does not become important until high temperature is
reached.
At the present stage in the evolution  of the Sun, the CNO
cycle is
believed
to play a rather small role in the energy and neutrino
production
budget (Bahcall 1989).\par
In the first reaction of the p-p chain, a proton decays into a
neutron in the immediate vicinity of another proton. The two
particles form a heavy variety of hydrogen known as deuterium,
along with a positron and an electronneutrino. There is a
second
reaction in the p-p chain producing deuterium and a neutrino
by
involving two protons and an electron. This reaction
(pep-reaction)
is 230 times less likely to occur in the solar core than the
first
reaction between two protons (pp-reaction). The deuterium
nucleus
produced in the pp- or pep-reaction fuses with another proton
to
form helium-3 and a gamma ray. About 88\% of the time the p-p
chain
is
completed when two helium-3 nuclei react to form an helium-4
nucleus and two protons, which may return to the beginning of
the
p-p chain. However, 12\% of the time, a helium-3 nucleus fuses
with
a helium-4 nucleus to produce beryllium-7 and a gamma ray. In
turn
the beryllium-7 nucleus absorbs an electron and transmutes
into
lithium-7 and an electronneutrino. Only once for every 5000
completions of the p-p chain, beryllium-7 reacts with a proton
to
produce boron-8 which immediately decays into two helium-4
nuclei,
a positron and an electronneutrino.\par
The net result of either the p-p chain or CNO cycle is the
production of helium nuclei and minor abundances of heavier
elements as $^7Be, ^7Li, ^8Be, ^8B$ (in the case of the p-p
chain)
or $^{13}N, ^{14}N, ^{15}N$ (in the case of the CNO cycle).
The
energy generated by thermonuclear reactions in the form of
gamma
rays is streaming (actually, diffusing) toward the solar
surface,
thereby getting scattered, absorbed and reemitted by nuclei
and
electrons. On their way outward, the high-energy gamma ray is
progressively changed to x-ray, to extreme ultraviolet ray, to
ultraviolet ray
and finally emerges mainly as visible light from the solar
surface
and radiates into
outer space. Only the weakly interacting neutrinos can leave
the
solar core with almost no interaction with solar matter.
However,
the chlorine experiment of Davis and collaborators, the
Japanese
Kamiokande experiment, and the GALLEX experiment at Gran Sasso
to
detect solar
neutrinos show that the Sun emits fewer of these elusive
particles
than the standard solar model predicts (Iben 1969; Lande 1989;
Hirata et al. 1991; Anselmann et al. 1992 a,b).
Since the beginning of the
70's this deficit challenges current understanding of solar
and
neutrino physics and of the process by which the Sun shines.
The
mystery of the missing solar neutrinos is commonly referred to
as
the ``solar neutrino problem''(Bahcall and Davis 1982).\par
\bigskip
\noindent
{\Large\bf Solar Evolution}\\

The general evolution scheme of the Sun postulates a
progressive
contraction of gas by self-gravitation which is periodically
interrupted by thermonuclear burning. After particular types
of
nuclear fuel (hydrogen, helium) are exhausted, the
contraction-burning cycle will be repeated, but at higher
temperatures. The
stages of the Sun's evolution from primitive solar nebula
contraction to the black-dwarf stage can be followed in the
Hertzsprung-Russell (H-R) diagram. There is a rapid movement
of the
Sun toward the main-sequence, where the Sun spends the major
part
of its life, and then an eventual movement toward the
black-dwarf
evolution stage, which is the final stage in its evolution
(Schwarzschild 1958; Gibson 1973).\par
\medskip
\noindent
{\large\bf Presolar evolution stages.}
\underline{cloud} Over 4.5 billion years ago, the gas cloud
which
would become
the
Sun had a diameter of over 480 trillion kilometers, which is
approximately 50 light years. This cloud was not dense,
containing
only a few thousand atoms per cubic centimeters of space. The
total
mass
of the cloud would have been sufficient for building up
several
solar systems. Its temperature was that of the interstellar
space,
of the order of 3 K, not radiating any light into the
surrounding space. The fragile equilibrium state of the cloud,
having only the choice of dissipating further into outer space
or
contracting into a denser configuration, eventually became
disturbed either by an impact from outside or by random
condensation of a
large number of cloud particles, and finally began to
condense.\\
\underline{globule}
After a time of the order of several thousand years, random
concentrations of matter called globules formed at various
places
in the giant condensing matter cloud. The cloud collapses
almost in
free fall, however, due to the influence of pressure, the
motion is
non homologous. The free fall time of the cloud is
\begin{equation}
t_{ff\;cloud} = (\frac{3\pi}{32G\rho_0})^{1/2},
\end{equation}
where $\rho_0$ denotes the initial mean matter density of the
cloud.
The temperature in the cloud
was
rising
very slowly, still not able to radiate light. Later, one
of those globules, now having a dimension of several hundred
solar systems, would become the Sun. The globule continued
condensing with the effect of increasing its temperature.\\
\underline{protostar}
Within 400,000 years the globule had condensed to a millionth
of
its
original size, but still over four times the size of the
present
solar
system.
At the centre of the globule a core had
developed, heated by the concentration of its matter, already
able
to radiate a substantial amount of energy into the less dense
outer
regions of the former globule. The emission of radiation by
the
core began
to slow the condensation of its matter. The matter becomes
opaque
and the free fall is stopped by the pressure. This core had
now
developed
into a stable and well-defined configuration called protostar
or
protosun. With the birth of the protosun the evolution of this
matter configuration advanced more rapidly. After the
formation of
a core, its free fall time is
\begin{equation}
t_{ff\;core} = (\frac{\pi^2}{8G}\frac{R^3}{M})^{1/2},
\end{equation}
where M is the mass and R the radius of the core,
respectively.
Within a few thousand
years
it collapsed to a size of the diameter of the orbit of planet
Mars.
The interior temperature reached values of 56,000K leading to
an
ionization
of atoms. The red light emitted at the surface of the protosun
was
not produced by fusion of atomic nuclei but by gravitational
contraction of matter. Gravitation released the potential
energy of
the globule, $7\times 10^{48} erg$, during the condensation of
the
protosun. According to the Virial theorem $(2T_k+\Omega=0)$
one
half
of the released gravitational energy $\Omega$ of the system
was
radiated from the
protosun while the other half had been transformed into heat
of the
central core; $T_k$ denotes the total kinetic energy of the
particles.\\
\underline{Sun}
Finally the protosun contracted further until its temperature
was
high enough for burning deuterium to form helium-3. The Sun
was
fully convective in the contraction phase and the chemical
composition was always uniform. Through deuterium burning the
contraction was momentarily slowed down. As the Sun continued
to
contract, the central temperature increased and the radiative
temperature gradient decreased relative to the convective
gradient.
Convection ceased and a radiative core grew outward. With the
ignition of hydrogen the protosun became a star,
characterized by the gravitationally stabilized fusion reactor
located at its center. Its binding gravitational energy
$\mid \Omega \mid\approx GM_\odot^2/R_\odot$ was initially
stored
in the extended globule,
called the primitive solar nebula. If the sun would shine by
its
store of thermal energy $\mid T_k\mid = \mid \Omega \mid /2$
(Virial theorem), then its lifetime is given by the Kelvin-
Helmholtz time scale, that is
\begin{equation}
t_{kw}=\frac{T_k}{L_\odot}\approx G\frac{M_\odot ^2}{R_\odot
L_\odot}.
\end{equation}
As the nuclear reactions began
to
release vast amounts of subatomic energy, the Sun was a
quite variable star,
varying in luminosity and surface activity as the result of
the
development of
a radiative core and convective currents in its outer layers
of
gas. After a period of some 30 million years, its structure
stabilized into the structure of a main-sequence star of one
solar
mass. The newly born Sun possessed enough
fuel in the form of hydrogen to keep shining steadily for a
time
period of the order of
\begin{equation}
t_{nuc}=\frac{E_{nuc}}{L_\odot}\approx 10^{-3}\frac{M_\odot
c^2}{L_\odot},
\end{equation}
where the factor  $10^{-3}$ is due to the product of the
percentage
of mass of the Sun available for hydrogen burning (0.1) and
the
fraction of mass converted into energy in hydrogen burning
$\Delta
m/4m_p\approx0.01$, where $\Delta m$ is the mass difference in
the net reaction $4p\rightarrow \alpha +2e^++2\nu_e$.\par
That means also that the
present Sun is right in the middle of its age as a
main-sequence
star (4.5 billion years).\\
{\large\bf Postsolar evolution stages.}
\underline{red giant} As the Sun ages, helium collects in its
center.
After a lifetime of 9 billion years as main-sequence star,
approximately 10\% of the hydrogen in the Sun's core will have
been
converted into helium and nuclear fusion reactions will cease
producing energy. The equilibrium between the total pressure
force
directed outwards and the gravitational force directed towards
the
centre of the Sun will be disturbed. The core of the Sun
starts
slowly collapsing under its own gravitational attraction.
Fusion
moves
outward to a shell surrounding the core, where hydrogen-rich
material is still present. The
gravitational
energy from the collapse will be converted into heat causing
the
shell to burn vigorously and so the
Sun's outer layers to swell immensely. The surface is now far
removed from the central energy source, cools and appears to
glow
red. The Sun now evolves into the
stage of a red giant. For a few hundred million years, the
expansion of the outer solar layers will continue, and the Sun
will
engulf the planet Mercury. The temperature on Venus and Earth
will
rise tremendously. Hydrogen fusion in the shell continues to
deposit helium ``ash'' onto the core, which becomes even
hotter and
more massive.\par
In the Sun's core nuclear fusion of helium into
carbon and oxygen will start to trigger even further the
expansion
of its outer layers.  The helium-rich core is unable to
lose heat fast enough and becomes unstable. In a very short
time of
few hours the core
gets too hot and is forced to expand explosively. Outer layers
of
the Sun will absorb the core explosion but the core will no
longer
be
able to produce energy by thermonuclear burning. Helium
fusion then continues in a shell and the structure of the Sun
would
look
like an onion: An outer, hydrogen-fusion layer and an inner,
helium-fusion layer which surrounds an inert core of carbon
and
oxygen.\par
The old
Sun may repeat the cycle of shrinking and swelling several
times.
In this stage of evolution the Sun is called an asymptotic
giant
branch star. Finally
enough carbon will accumulate in the core to prevent the
core explosion. Helium-shell burning will add heat to the
outer
layers of the Sun, mainly containing hydrogen and helium. The
asymptotic giant Sun will generate eventually an intense wind
that
begins to carry off its outer envelope. The precise mechanism
behind this
phenomenon is not yet well understood. The Sun
will expand a final time and after about 30 million years it
will
swallow Venus and Earth, outer layers will keep expanding
outward
and as much as half of the Sun's mass gets lost into space.\\
\underline{white dwarf}
The solar core keeps shrinking and because it is not able
anymore
to produce radiation by
fusion the further evolution of this configuration is governed
by
gravitation. All matter will collapse into a small body about
the
size
of the Earth. Thus, the Sun will have become a white dwarf,
this is
a dense-matter configuration, having radiated away the energy
of
its
collapse. Then the white dwarf rapidly begins
to
cool.\\
\underline{black dwarf}
The final stage of solar evolution will be the black dwarf
stage.
The white dwarf will emit yellow light and then red light in
the
course of its evolution, drawing from the star's reservoir of
thermal energy. Its nuclei will be packed as tightly as
physically possible and no further collapse is possible. The
body
is
progressively cooling down and finally becomes as cold as the
interstellar space
around it, emitting no light at all. As a carbon-oxygen-rich
black
dwarf
it will continue its journey through the galaxy (milky way)
and may eventually encounter another giant gas cloud to become
involved in
the birth of a new star.\par

\bigskip
\begin{flushright}
HANS J. HAUBOLD\par
\medskip
A.M. MATHAI
\end{flushright}
\clearpage
\begin{center}
{\Large\bf References}
\end{center}
ALLEN, C.W. 1973. \underline{Astrophysical Quantities}.
London:\par
The Athlone Press.\par
\noindent
ANSELMANN, P. et al. 1992a. Solar neutrinos observed by GALLEX
at\par
Gran Sasso. Physics Letters B285: 376-389.\par
\noindent
ANSELMANN, P. et al. 1992b. Implications of the GALLEX
determination\par
of the solar neutrino flux. Physics Letters B285: 390-397.\par
\noindent
BAHCALL, J.N., and R. DAVIS Jr., 1982. An account of the
development\par
of the solar neutrino problem. In \underline{Essays in
Nuclear Astrophysics}, ed.\par
C.A. Barnes, D.D. Clayton, and D.N.
Schramm, pp. 243-285. Cambridge:\par
Cambridge University Press.\par
\noindent
BAHCALL, J.N. 1989. \underline{Neutrino Astrophysics}.
Cambridge:
Cambridge\par
University Press.\par
\noindent
BEER, J. 1987. Radioisotopes in natural archives:
information about the\par
history of the solar-terrestrial system.
In \underline{Solar-Terrestrial}\par
\underline{ Relationships and the Earth
Environment in the Last Millennia},\par
Proceedings of the International School of Physics
$\ll$ Enrico Fermi $\gg$,\par
Course XCV, ed. G. Cini, pp. 183-198.
Amsterdam: North Holland.\par
\noindent
BRACEWELL, R.N. 1989. The solar cycle: A central-source wave
theory.\par
\underline{Proceedings of the Astronomical Society
of Australia} 8(2):145-147.\par
\noindent
CHANDRASEKHAR, S. 1967. \underline{An Introduction to the
Study of
Stellar}\par
\underline{Structure}.
New York: Dover Publications, Inc.\par
\clearpage
\noindent
DEUBNER, F.-L., and D. GOUGH. 1984. Helioseismology:
oscillations
as a\par
diagnostic of the solar interior. \underline{Annual Review of
Astronomy and}\par
\underline{Astrophysics} 22: 593-619.\par
\noindent
DICKE, R.H., and H.M. GOLDENBERG. 1967. Solar oblateness and
general relativity. \underline{Physical Review Letters}
18:313-316.\par
\noindent
DICKE, R.H., J.R. KUHN, and K.G. LIBBRECHT. 1987.\par
Is the solar oblateness variable? Measurements of 1985.\par
\underline{The Astrophysical
Journal} 318:451-458.\par
\noindent
FAIRBRIDGE, R.W. 1987a. Sunspots. In \underline{The
Encyclopedia
of Climatology},\par
eds. J.E. Oliver and R.W. Fairbridge, pp. 815-823.
New York:\par
Van Nostrand Reinhold Company.\par
\noindent
FAIRBRIDGE, R.W. 1987b. Climatic variation, historical record.
In\par
\underline{The Encyclopedia of Climatology}, eds. J.E. Oliver
and\par
R.W. Fairbridge, pp. 305-323.
New York:\par
Van Nostrand Reinhold Company.\par
\noindent
GIBSON, E.G. 1973. \underline{The Quiet Sun}. NASA SP-303.\par
\noindent
GUENTHER, D.B., P. DEMARQUE, Y.-C. KIM, and M.H.
PINSONNEAULT.\par
1992. Standard solar model. \underline{The Astrophysical
Journal}
387:372-393.
\par
\noindent
HAUBOLD, H.J., and A.M. MATHAI. 1987. Analytical results\par
connecting stellar structure parameters and neutrino
fluxes.\par
\underline{Annalen der Physik (Leipzig)} 44(2):103-116.\par
\clearpage
\noindent
HAUBOLD, H.J., and A.M. MATHAI. 1992. Analytic stellar
structure.\par
\underline{Astrophysics and Space Science} 197:153-161.\par
\noindent
HERMAN, J.R., and R.A. GOLDBERG. 1978. \underline{Sun,
Weather, and
Climate}.\par
NASA-SP-426.\par
\noindent
HILL, H.A., and R.T. STEBBINS. 1975. The intrinsic visual
oblateness\par
of the Sun.\underline{The Astrophysical
Journal} 200: 471.\par
\noindent
HILL, H.A., and R.J. KROLL. 1992. Long-term solar variability
and
solar\par
seismology: I. In \underline{Basic Space Science}, eds. H.J.
Haubold,
and R.K. Khanna,\par
pp. 170-180. New York: American Institute
of Physics
Conference\par
Proceedings Vol. 245.\par
\noindent
HILL, H.A., P. OGLESBY, and Y.-M. GU. 1992. Long-term solar
variability\par
and solar seismology: II. In \underline{Basic Space
Science}, eds. H.J. Haubold, and\par
R.K. Khanna, pp. 181-192. New York:
American Institute of Physics\par
Conference Proceedings Vol. 245.\par
\noindent
HIRATA, K.S. et al. 1991. Real-time, directional
measurement\par
of $^8B$ solar neutrinos in the Kamiokande-II detector.
Physical\par
Review D44: 2241-2260.\par
\noindent
IBEN Jr., I. 1969. The $Cl^{37}$ solar neutrino experiment and
the
solar helium\par
abundance. \underline{Annals of Physics} 54 (1):164-203.\par
\noindent
KAVANAGH, R.W. 1972. Reaction rates in the proton-proton
chain.\par
In \underline{Cosmology, Fusion and Other Matters}, ed. F.
Reines,
pp.
169-185.\par
Boulder: Colorado Associated University Press.\par
\noindent
KOURGANOFF, V. 1973. \underline{Introduction to the Physics of
Stellar Interiors}.\par
Dordrecht: D. Reidel Publishing Company.\par
\noindent
KUNDT, W. 1992. The 22-year cycle of the Sun.
\underline{Astrophysics and Space Science}\par
187:75-85.\par
\noindent
LANDE, K. 1989. Status of solar neutrino observations and
prospects
for\par
future experiments. \underline{Annals of the New York Academy
of Sciences}\par
571:553-560.\par
\noindent
LANG, K.R. 1980. \underline{Astrophysical Formulae.} New York:
Springer-Verlag.\par
\noindent
LONGAIR, M. 1989. The new astrophysics. In \underline{The New
Physics},\par
ed. P. Davies, pp. 94-208. Cambridge: Cambridge
University Press.\par
\noindent
MATHAI, A.M., and H.J. HAUBOLD. 1986. \underline{Modern
Problems in
Nuclear and}\par
\underline{Neutrino Astrophysics}. Berlin:
Akademie-Verlag.\par
\noindent
SACKMANN, I.-J., A.I. BOOTHRAYD, and W.A. FOWLER. 1990. Our
Sun.I.\par
The standard model: successes and failures. \underline{The
Astrophysical Journal}\par
360:727-736.\par
\noindent
SCHATTEN, K.H., and A. ARKING (eds.). 1990. \underline{Climate
Impact of Solar}\par
\underline{Variability}. NASA-CP-3086.\par
\noindent
SCHOU, J., J. CHRISTENSEN-DALSGAARD, and M.J. THOMPSON.
1992.\par
The resolving power of current helioseismic inversions
for the Sun's\par
internal rotation.
\underline{The Astrophysical Journal} 385:L59-L62.\par
\noindent
SCHWARZSCHILD, M. 1958. \underline{Structure and Evolution of
the
Stars.}\par
New York: Dover Publications, Inc. \par
\noindent
SEARS, R.L. 1964. Helium content and neutrino fluxes.\par
\underline{The Astrophysical Journal}
140:477-484.\par
\noindent
SOFIA, S. (ed.). 1981. \underline{Variations of the Solar
Constant.} NASA-CP-2191.\par
\noindent
SONETT, C.P. 1984. Very long solar periods and the radiocarbon
record.\par
\underline{Reviews of Geophysics and Space Physics},
22(3):239-254.\par
\noindent
STIX, M. 1989. \underline{The Sun.} Berlin:
Springer-Verlag.\par
\noindent
WEISS, W.W., and H. SCHNEIDER. 1991. Astroseismology.\par
\underline{The Messenger} 66 (December):36-40.\par
\noindent
ZIRIN, H. 1988. \underline{Astrophysics of the Sun.}
Cambridge:
Cambridge University\par
Press.
\clearpage
\noindent
TABLE CAPTIONS\par
\bigskip
\begin{tabbing}
Table 1-1. \hspace{0.5cm}\=A number of physical
characteristics of
the Sun.\\
[1cm]
Table 1-2. \>Internal structure of the Sun
$(R_\odot=6.96\times10^5\; km)$\\ [1cm]
Table 1-3. \>Principal reactions of the proton-proton-chain\\
\>and the CNO cycle in the Sun.
\end{tabbing}
\clearpage
\noindent
FIGURE CAPTIONS\par
\bigskip
\begin{tabbing}
Figure 1-1. \hspace{0.5cm}\=The beauty of the Sun. This
snapshot\\
\>from NASA's Skylap 4 (19 December 1973) shows one of the\\
\>most spectacular solar flares ever recorded, spanning more\\
\>than 588,000 km across the solar surface (Courtesy NASA\\
\>Goddard Space Flight Center).\\ [0.5cm]
Figure 1-2. \>A standard solar model of the present solar\\
\>interior: $X=0.708, Y=0.272, Z=0.0020, \rho_c=158
gcm^{-3}$,\\
\>$T_c=1.57\times10^7 K.$ (Courtesy of R.L. Sears of
University\\
\>of Michigan; Sears 1964).\\ [0.5cm]
Figure 1-3.\>Schematic view of the structure of the Sun and
modes\\
\>of outward flow of energy (Courtesy NASA Goddard Space\\
\>Flight Center).\\ [0.5cm]
Figur 1-4.\>Solar interior rotation profile, as inferred from
an\\
\>inversion of p-mode splitting data, displayed at these\\
\>latitudes: $\Theta_0=0^\circ$ (polar), $\Theta_0=45^\circ$,
and $\Theta_0=90^\circ$\\
\>(equatorial). Dashed lines indicate $1\sigma$ error bars
based\\
\>on the observer's estimates of the uncertainties in the\\
\>measured $a_j$ coefficients (Courtesy J.
Christensen-Dalsgaard;\\
\>Schou et al. 1992).\\ [0.5cm]
Figure 1-5.\> Solar p-modes, equatorial cross section\\
\>$l=40, m=40, \nu=3.175 mHz$ (Courtesy W.W. Weiss; \\
\>Weiss and Schneider 1991).\\ [0.5cm]
Figure 1-6.\>Solar p-modes; equatorial cross section\\
\>$l=40, m=0, \nu=3.175 mHz$ (Courtesy W.W. Weiss; Weiss and\\
\>Schneider 1991).\\ [0.5cm]
Figure 1-7. \>Solar p-modes, equatorial cross section\\
\>$l=2, m=2, \nu=3.147 mHz$ (Courtesy W.W. Weiss; Weiss and\\
\>Schneider 1991).\\ [0.5cm]
Figure 1-8.\>The path of a $1M_\odot$ star in the
Hertzsprung-\\
\>Russel diagram.
\end{tabbing}
\clearpage
\noindent
TABLE 1-1 A number of physical characteristics of the Sun\par
\bigskip
\noindent
\begin{tabbing}
\=Mean distance from Earth \hspace{2cm}\=1 astronomical unit
(A.U.)=\\
\>\>$1.496 \times 10^8 km$\\
\>Radius \>$R_\odot = 6.960 \times 10^5 km$\\
\>Mass \>$M_\odot = 1.991 \times 10^{33} g$\\
\>Mean density \>$\bar{\rho}_\odot = 1.410 gcm^{-3}$\\
\>Gravity at surface \>$g_\odot = 2.738 \times 10^4
cms^{-2}$\\
\>Total energy output (luminosity) \>$L_\odot = 3.860 \times
10^{33}
ergs^{-1}$\\
\>Effective surface temperature \>$T_{eff} = 5780 K$\\
\>Solar age \>$t_\odot = 4.5 \times 10^9 yr$
\end{tabbing}
\clearpage
TABLE 1-2 Internal structure of the Sun
$(R_\odot=6.96\times10^5
km)$\par
\bigskip
\begin{tabular}{lcl}
Internal region & extension in terms & chemical composition\\
& of solar radius &\\ [0.5cm] \hline
core & $0.20 R_\odot$ & center only:\\
& & He: 0.63, H: 0.35,\\
& & metals: 0.02 \\
& & (almost actively ionized matter)\\ [0.3cm]
radiative zone & $0.50 R_\odot$ & He: 0.23, H: 0.75,\\
& & metals: 0.02\\
& & (highly ionized)\\ [0.3cm]
convective zone & $0.30 R_\odot$ & same (less ionized)\\
[0.3cm]
photosphere & $0.002 R_\odot$ & same (less ionized)\\ [0.3cm]
solar surface & $1.000 R_\odot$ &\\ [0.5cm]\hline
chromosphere & $0.02$ & same (less ionized)\\ [0.3cm]
corona & $\approx 5$ & same (highly ionized)\\
\end{tabular}
\clearpage
\renewcommand{\baselinestretch}{1}
\small\normalsize
TABLE 1-3 Principal reactions of the proton-proton chain and
the
CNO\par
cycle in the Sun\par
\vspace{0.5cm}
\begin{tabular}{llcl}
Number & Reaction & Termination & Neutrino Energy\\
& &(\%) & (MeV)\\ \hline
p-p chain & & \\ [0.3cm]
1 & $p+p \longrightarrow ^2H+e^++\nu_e$ & $99.75$ & $0.420$
(spectrum)\\ [0.3cm]
& or & & \\ [0.3cm]
2 & $p+e^-+p \longrightarrow ^2H+\nu_e$ & $0.25$ & $1.44$
(line)\\
[0.3cm]
3 & $^2H+p \longrightarrow ^3He+\gamma$ & $100$ & \\ [0.3cm]
4 & $^3He+^3He \longrightarrow ^4He+2p$ & $88$ & \\ [0.3cm]
& or & & \\ [0.3cm]
5 & $^3He+^4He \longrightarrow ^7Be+\gamma$ & $12$ & \\
[0.3cm]
6 & $^7Be+e^- \longrightarrow ^7Li+\nu_e$ & $99.98$ & $0.861
(90\%)$\\ [0.3cm]
& & & $0.383 (10\%)$\\ [0.3cm]
& & & (both lines)\\ [0.3cm]
7 & $^7Li+p \longrightarrow 2^4He$ & & \\ [0.3cm]
& or & & \\ [0.3cm]
8 & $^7Be+p \longrightarrow ^8B+\gamma$ & $0.02$ & \\ [0.3cm]
9 & $^8B \longrightarrow ^8Be^*+e^++\nu_e$ & & $14.06$
(spectrum)\\
[0.3cm]
10 & $^8Be^* \longrightarrow 2^4He$ & & \\ [0.3cm]
CNO cycle & & \\ [0.3cm]
1 & $^{12}C+^1H \longrightarrow ^{13}N+\gamma$ & & \\ [0.3cm]
2 & $^{13}N \longrightarrow ^{13}C+e^++\nu_e$ & & $1.2$
(spectrum)
\\ [0.3cm]
3 & $^{13}C+^1H \longrightarrow ^{14}N+\gamma$ & & \\ [0.3cm]
4 & $^{14}N+^1H \longrightarrow ^{15}O+\gamma$ & & \\ [0.3cm]
5 & $^{15}O \longrightarrow ^{15}N+e^++\nu_e$ & & $1.7$
(spectrum)
\\ [0.3cm]
6 & $^{15}N+^1H \longrightarrow ^{12}C+^4He$ & & \\
\end{tabular}
\end{document}